\documentclass{aa}

\usepackage{graphics}
\usepackage{longtable}
\usepackage{txfonts}
\usepackage{graphicx}
\usepackage{hyperref}
\begin{document}

\title{VLTI/MIDI observations of 7 classical Be stars\thanks{Based on observations made with ESO Telescopes at Paranal under  programs 073.A-9014G, 077.D-0097 and 082.D-0189}}

\authorrunning{Meilland et al.}
\titlerunning{VLTI/MIDI observations of 7 classical Be stars}

\author{ A. Meilland \inst{1}, Ph. Stee \inst{2}, O. Chesneau \inst{2}, and C. Jones \inst{3}}

   \offprints{meilland@mpifr-bonn.mpg.de}

\institute{
Max Planck Institut fur Radioastronomie, Auf dem Hugel 69, 53121 Bonn, Germany 
\and
UMR 6525 CNRS FIZEAU UNS, OCA, CNRS, Campus Valrose, F-06108 Nice cedex 2, France.
\and
Physics and Astronomy Department, The University of Western Ontario, London, Ontario, Canada, N6A 3K7.
}

   \date{Received; accepted }

   \abstract{Classical Be stars are hot non-supergiant stars surrounded by a gaseous circumstellar envelope that is responsible for the observed IR-excess and emission lines.  The origin of this envelope, its geometry, and kinematics have been debated for some time.}
{We measured the mid-infrared extension of the gaseous disk surrounding seven of the closest Be stars in order to constrain the geometry of their circumstellar environments and to try to infer physical parameters characterizing these disks.}
{Long baseline interferometry is the only technique that enables spatial resolution of the circumstellar environment of classical Be stars. We used the VLTI/MIDI instrument with baselines up to 130 m  to obtain an
angular resolution of about 15~mas in the N band and compared our results with previous K band measurements obtained with the VLTI/AMBER instrument and/or the CHARA interferometer.}
{We obtained one calibrated visibility measurement for each of the four stars, p Car, $\zeta$ Tau, $\kappa$ CMa, and $\alpha$ Col, two for $\delta$ Cen and $\beta$ CMi, and three for $\alpha$ Ara. Almost all targets remain unresolved even with the largest VLTI baseline of 130m, evidence that their circumstellar disk extension is less than 10~mas. The only exception is $\alpha$ Ara, which is clearly resolved and well-fitted by an elliptical envelope with a major axis a=5.8$\pm$0.8~mas and an axis ratio a/b=2.4$\pm$1 at 8 $\mu$m. This extension is  similar to the size and flattening measured with the VLTI/AMBER instrument in the K band at 2 $\mu$m.}
{The size of the circumstellar envelopes for these classical Be stars does not seem to vary strongly on the observed wavelength between 8 and 12 $\mu$m. Moreover, the size and shape of $\alpha$ Ara's disk is almost identical at 2, 8, and 12 $\mu$m. We expected that longer wavelengths probe cooler regions and correspondingly larger envelopes, but this is clearly not the case from these measurements. For $\alpha$ Ara this could come from to disk truncation by a small companion ; however, other explanations are needed for the other targets.}


   \maketitle
%

\section{Introduction} 

Classical Be stars are hot non-supergiant stars that are surrounded by a gaseous circumstellar environment responsible for the presence of an IR-excess and emission lines in their spectra. A generally accepted view of this material is a dense equatorial region dominated by rotation, often called a "disk" where most of the IR-excess and emission lines originate,
surrounded by a more diluted, radiatively driven, polar wind. Despite the
fact that these systems have been studied for decades, the disk and wind formation, geometry, and kinematics, remain contentious issues.

First VLTI observations of 4 Be stars, namely Achernar (Domiciano de Souza et al. 2003), $\alpha$ Ara (Meilland et al. 2007a), $\kappa$ CMa (Meilland et al. 2007b), and $\delta$ Cen (Meilland et al. 2008) have shown evidence of different geometries likely to due different mass-ejection processes such as fast rotation, radiatively driven winds or binarity. Thus, it is not clear if  Be stars can be considered as a homogeneous group of stars in terms of mass-ejection processes (Stee \& Meilland 2009).

Measuring the N-band extension of the circumstellar environments of classical Be stars can put strong constraints on their physical parameters as shown by Chesneau et al. (2005) for $\alpha$ Ara. For this star, VLTI/MIDI observations with 100~m baselines led to nearly unresolved visibilities, demonstrating that the envelope was even smaller than what was previously predicted by Stee \& Bittar (2001). These authors used long term spectroscopic variability measurements and concluded that the disk is likely truncated by a companion.

We initiated an observational campaign on the brightest, closest, southern hemisphere classical Be stars using the VLTI instruments MIDI and AMBER. In this paper we present new spectrally resolved mid-infrared VLTI/MIDI interferometric measurements for seven classical Be stars, p Car, $\zeta$~Tau, $\beta$ CMi, $\kappa$ CMa, $\alpha$ Col, $\delta$ Cen, and $\alpha$ Ara. 

The paper is organized as follows. In Sect. 2 we present the observations and the data reduction process. In Sect. 3, the calibrated visibilities are presented and then compared with a model composed of a unresolved star plus uniform disk. Our results and conclusions are discussed in Sect. 4, and Sect. 5, respectively.

\section{Observations and data reduction}
VLTI/MIDI observations for our program stars were carried out at Paranal Observatory between June 2004 and January 2009 with the 8.2 m UT telescopes. We obtained one visibility measurement for p Car, $\zeta$~Tau, $\kappa$ CMa, and $\alpha$ Col, two for $\delta$ Cen and $\beta$ CMi, and three for $\alpha$ Ara. Observations of 12 calibrator stars were also
obtained during this campaign: HD40808, HD49161, HD50778, HD94452, HD 95272, HD107446, HD151249, HD163376, HD169916, HD198048, HD217902, and HD218594. The observation log is presented in Table 1.

The 2004 observations of $\delta$ Cen were obtained with UT1-UT3 in HIGH-SENS mode. All of the other observations were taken with the UT1-UT4 baseline using the SCI-PHOT mode to enable a better visibility calibration (Chesneau~\cite{chesneau07}). The PRISM low-spectral dispersion mode, allowed us to obtain spectrally resolved visibilities with R=30 across the N band. We have used the standard MIA + EWS, (version 1.6) package to reduce the data and estimate the errors bars  (see Fig.~\ref{visi}).

\begin{table}[!t]
{\centering \begin{tabular}{@{}c@{~~}c@{~~}c@{~~}c@{~~}c@{~~}c@{~~}c@{~~}c@{}}
\hline Star & Obs. Date & Time & \multicolumn{3}{c}{Baseline} & Calibrators\\
&&(UT)&B$_n$&L(m)& P.A.($^o$)&(HD)\\
\hline
$\kappa$ CMa&2008/12/14&02:33&&117.0&25.2&40808, 94452\\
$\delta$ Cen&2004/06/06&00:43&B$_2$&91.2&42.5&107446\\
						&2008/12/13&08:27&B$_1$&129.2&27.7&40808, 94452\\
$\alpha$ Ara&2006/04/15&03:50&B$_1$&126.8&1.7&95272, 129456, 151249\\
						&2006/04/16&07:49&B$_2$&129.6&54.4&151249, 163376\\
						&2006/07/08&04:15&B$_3$&120.7&74.1&169916, 198048, 217902\\
						&					 &		 & &		 &		& 218594\\
p Car				&2008/12/13&07:44&&129.5&40.8&40808, 94452\\
$\zeta$ Tau	&2009/01/11&03:36&&127.3&67.9&49161, 50778\\
$\alpha$ Col&2008/12/13&05:17&&128.5&64.0&40808, 94452\\
$\beta$ CMi	&2008/12/17&07:07&B$_1$&129.5&63.8&49161, 50778\\
						&2009/01/11&04:44&B$_2$&125.3&64.5&49161, 50778\\
\hline
\end{tabular}\par}
\label{MIDI_log}
\caption{VLTI/MIDI observing log for our program stars and their corresponding calibrators.}
\end{table}

\section{Results}

\subsection{The calibrated visibilities}

\begin{figure}[!ht]
\centering   \includegraphics[width=0.495\textwidth]{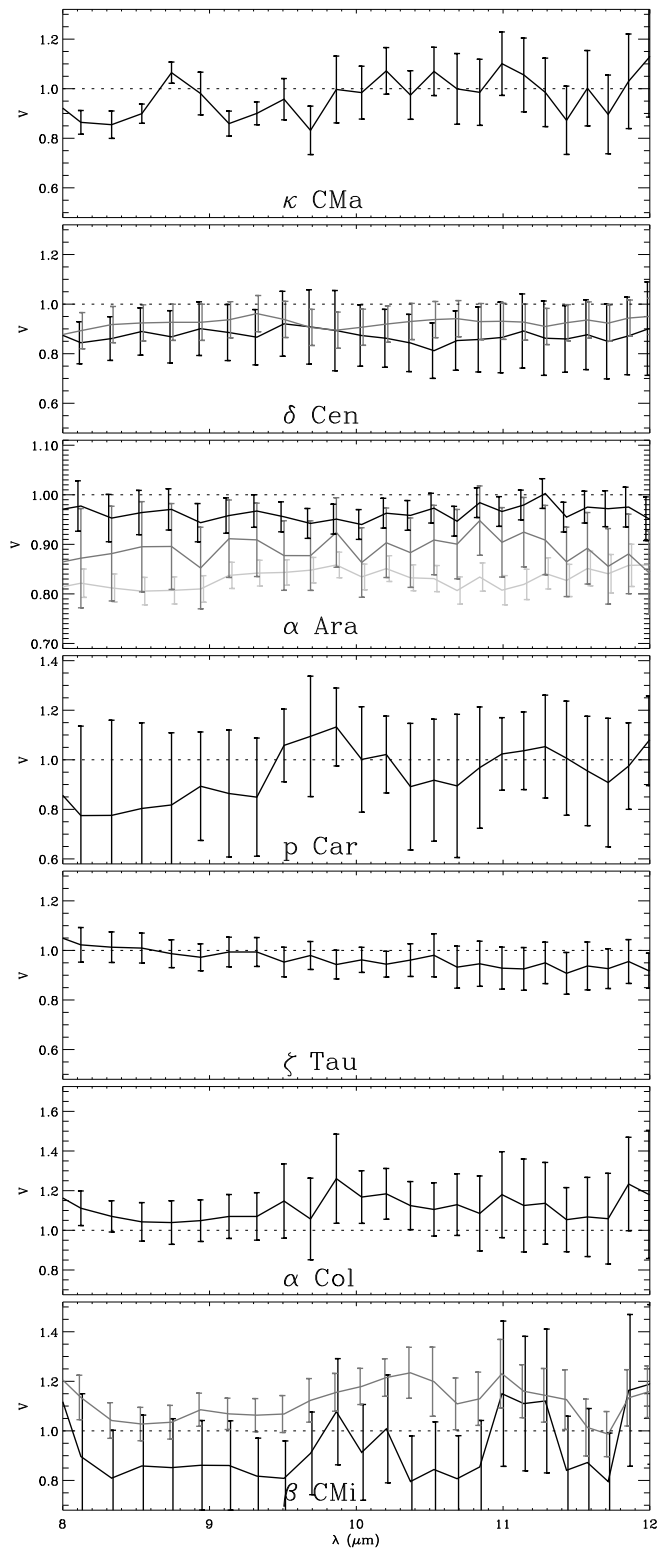}
\caption{VLTI/MIDI visibilities of the 7 Be stars plotted as a function of wavelength across the N band. For stars with multiple measurements (i.e., $\alpha$ Ara $\beta$ CMi, and $\delta$ Cen) the corresponding baselines B$_1$, B$_2$, and B$_3$ are plotted in black, gray, and light gray, respectively. See Table 1 for description of the baselines.}
\label{visi}
\end{figure}

The N band calibrated visibilities for the program Be stars are plotted as a function of wavelength in Fig~\ref{visi}. The only clearly
resolved (i.e., V+3$\sigma<$1) target is $\alpha$ Ara along the B$_1$ baseline. It is also barely resolved (i.e., V+$\sigma<$1) along the 
other two baselines. Moreover, $\delta$ Cen (for both baselines) 
and $\zeta$ Tau (for $\lambda>$10$\mu$m) are also barely resolved. All of 
the other targets,  $\kappa$ CMa, p Car, $\alpha$ Col, and $\beta$ CMi, are clearly unresolved. Therefore, when we take into account the uncertainty 
in the measurements for these targets, we can only determine an upper limit of the N band envelope extension.

All targets have visibilities larger than 0.8. This corresponds, for a 130m baseline 
at 8$\mu$m to extensions smaller than 6~mas assuming an uniform disk or 
3~mas assuming a FWHM Gaussian. However, to estimate an accurate envelope extension, we need to include the contribution from the central star to the total flux (star+envelope) for each of systems studied.

\begin{table}[!b]

{\centering \begin{tabular}{cccc}
\hline
Name&Distance (pc)& Spectral Class& $\Phi_\star$ (mas)\\
\hline
$\kappa$ CMa& 202 $\pm$ 5& B1.5IV&0.44 $\pm$ 0.15\\
$\delta$ Cen&127 $\pm$ 8& B2IV&0.65 $\pm$ 0.3\\
$\alpha$ Ara&82 $\pm$ 6& B3V&0.88 $\pm$ 0.2\\
p Car&148 $\pm$ 9& B4V&0.27$ \pm$ 0.15\\
$\zeta$ Tau&136 $\pm$ 15& B4III&0.65 $\pm$ 0.2\\
$\alpha$ Col&80 $\pm$ 2& B7IV&0.58 $\pm$ 0.15\\
$\beta$ CMi&50 $\pm$ 1& B8V&0.56 $\pm$ 0.1\\
\hline
\end{tabular}\par}

\caption{Distance, spectral class, and estimated stellar angular diameters ($\Phi_\star$)\label{dist_sptype} for the seven observed Be stars.}
\end{table}

\begin{table*}[t]
{\centering \begin{tabular}{|c|c|c|ccccccc|}
\hline
Reference &Method &Effect of &\multicolumn{7}{c}{Effective Temperature [K]}\vline\\
& & Rotation & $\kappa$ CMa& $\delta$ Cen& $\alpha$ Ara& p Car&$\zeta$ Tau& $\alpha$ Col& $\beta$ CMi\\
\hline
Dachs et al. (1988)         &Spectro-Photometry  &No &22500&22500&18000&16500&-    &13000&-\\
Dachs et al. (1990)         &Balmer Decrement&No &22500&20000&18000&22500&-    &-&-\\
Chauville et al. (2001)     &Spectroscopy&No &21330&21527&19010&17906&20137&13031&-    \\
Fremat et al. (2005)        &Spectroscopy&Yes&24627&22360&18044&17389&19310&12963&11772\\
Levenhagen \& Leister (2006)&Spectroscopy&No &24100&22230&22150&-    &-    &14200&13100\\
\hline
\end{tabular}\par}
\caption{Published effective temperature for our target stars.\label{teff}}
\end{table*}

\subsection{A simple two component model}

If we assume, for all our targets,  that the N band emission originates from the central star and and the circumstellar envelope only, we can write~:

\begin{equation}
\rm {V_{ tot}=V_eF_e + V_\star (1-F_e)},
\end{equation}

where V$_{\rm tot}$, V$_{\rm \star}$, and V$_{\rm e}$, are the total, stellar, and envelope visibilities, respectively, and F$_e$, is the ratio of the envelope flux to the total object flux. 

We can estimate the stellar angular diameter using the spectral class of each object and their distance derived from Hipparcos parallax measurements (\textit{Hipparcos, the New Reduction}  van Leeuwen, 2007). As shown in Table~\ref{dist_sptype}, all targets have stellar diameters smaller than 1~mas. Thus, V$_{\rm \star}>$ 0.99 at 8$\mu$m and we can assume V$_{\rm \star}$=1 in the entire N band for these stars. 

Using this assumption, we can rewrite Eq.(1) :

\begin{equation}
\rm {V_e=\frac{V_{tot}-1+F_e}{F_e}}.
\end{equation}

Consequently, in order to calculate the envelope visibility V$_{\rm e}$ we only need the relative envelope flux F$_{\rm e}$.

\subsection{Reconstructing the spectral energy distribution}

\begin{figure}[!ht]
\centering   \includegraphics[width=0.49\textwidth]{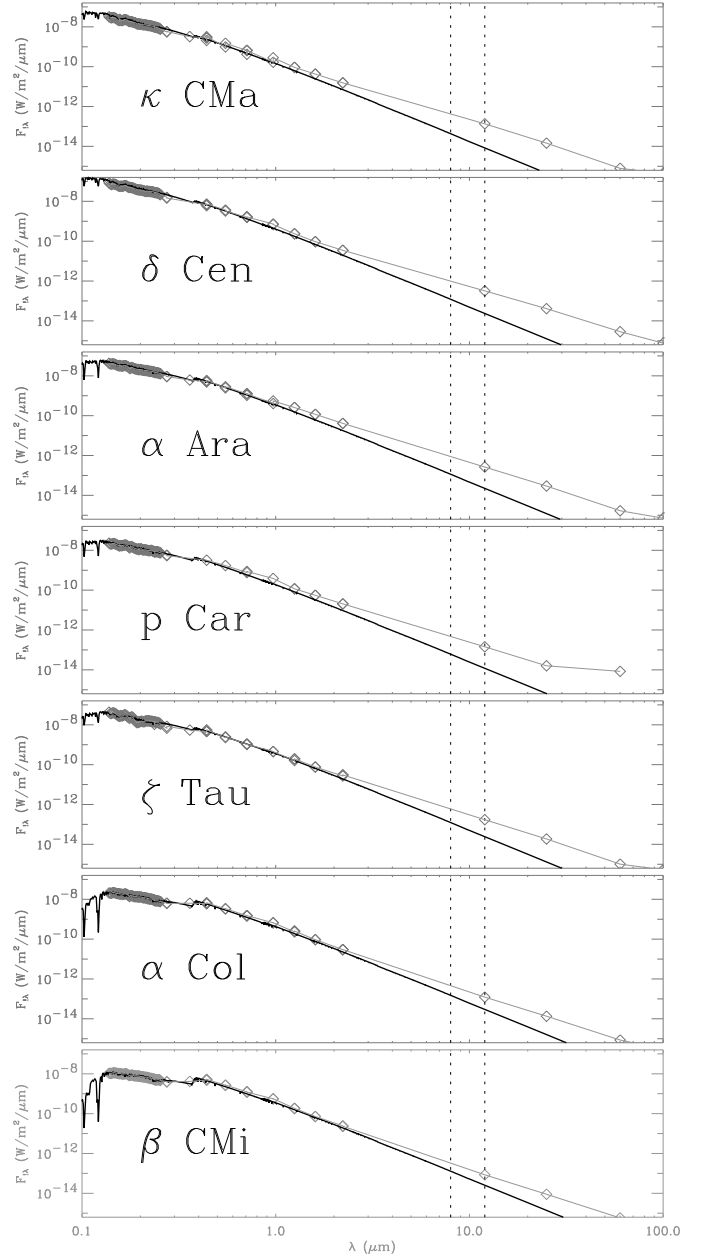}
\caption{Reconstructed SEDs for the 7 Be star systems observed with the VLTI/MIDI instrument are shown with gray solid lines. Diamonds represent measurements taken from the VIZIER database. The portions of the SEDs up to 1$\mu$m (solid black lines) are fitted using reddened Kurucz models with the stellar parameters given in Table 4. Dashed lines represent the wavelengths for which the envelope flux is calculated, i.e., at 8$\mu$m, and 12$\mu$m. }
\label{SED}
\end{figure}

We chose to constrain F$_{\rm e}$ by fitting the objects' spectral energy distribution (SED) using stellar models. For each target, we reconstructed its SED using data available from the VIZIER\footnote{http://vizier.u-strasbg.fr/viz-bin/VizieR} database (see Fig~\ref{SED}): UV  measurements are from Jamar et al. (1976) and Thompson et al. (1978),  magnitudes are from Morel \& Magnenat (1978), Monet et al. (2003), Zacharias et al. (2005), and 2MASS (Cutri et al., 2003),  and IRAS measurements are used for the mid to far infrared. We note that Be stars are observed to exhibit photometric variations with timescales from a few months to several years with amplitudes up to 0.5 Mag (Harmanec, 1983) so that the SEDs can slightly vary with time. However, contemporaneous UV to mid-IR photometry 
are not available to compare with our VLTI/MIDI measurements so we use 
measurements from 0.2 to 100$\mu$m obtained at other times. The resulting reconstructed SEDs for our targets are plotted in Fig~\ref{SED}.

\subsection{The stellar parameters and distance}

In order to constrain the stellar contribution to the flux we need 
T$_{\rm eff}$, R$_{\rm \star}$,  and g$_{\rm eff}$ for the central star. 
However, these stellar parameters for rapidly rotating Be stars are quite uncertain. There are several methods in the literature that are
used to determine these stellar parameters. For example, Dachs et al. (1989) derived stellar parameters and extinctions of 46 Be stars from UBV and JHKLMN photometric measurements. Later, Dachs et al. (1990) determined the stellar parameters of 26 objects from an analysis of the Balmer decrement. Chauville et al. (2001) and Levenhagen \& Leister (2006) use spectroscopic measurements of photospheric lines to calculate these parameters. Table \ref{teff} summarizes T$_{\rm eff}$ derived by all of these authors for our target stars. 

The discrepancy between T$_{\rm eff}$ determined with these methods for classical Be stars can reach several thousand of Kelvin for some targets (see $\zeta$ Tau for example). These ranges mainly stems from the latitudinal-dependent surface temperatures induced by rapid rotation. 
This effect will depend on the inclination angle of the star. Also, the circumstellar gas can cause reddening and 
obscure the star affecting the values of the spectrophotometrically determined parameters.

Fremat et al. (2005) accounted rapid rotation in their analyzes and quoted ``apparent" stellar T$_{\rm eff}$ and g$_{\rm eff}$, and then compared these values to non-rotating counter-parts that are typically 
thought to be $\sim$ 5 to 15$\%$ hotter. We adopt these apparent or average values for our work. The apparent T$_{\rm eff}$ derived by Fremat et al. (2005) for our sample stars are also presented in Table~\ref{teff}. 

Assigning a value for the stellar radius for a rapidly rotating star can also
be difficult since the polar radius and equatorial radius can be quite different.
For our targets, typical values for the stellar radius are deduced from the objects' spectral class and then adjusted to obtain the best SED fit in the UV and visible. The fit of the 
stellar flux is accomplished using Kurucz models (Kurucz, 1979). We note that the use of 
plan parallel atmospheric models with a global T$_{\rm eff}$ for fast rotating star may not 
apply. However, the effect on the mid-infrared flux should be negligible for these 
wavelengths since the stellar radiation is dominated by black-body emission. The Kurucz models are reddened using the law of extinction from Cardelli et al.(1989) combined with the standard interstellar 
R$_{\rm v}$ value of 3.1. Thus, with T$_{\rm eff}$ and log g taken from Fremat et al. (2005), and 
the distance derived from Hipparcos parallaxes measurements, only the values for 
R$_{\rm \star}$ and A$_{\rm v}$ are determined from the fit of the SED. 

\subsection{Estimating the errors on the envelope relative flux}

To estimate the errors in the relative envelope flux F$_{\rm e}$ at 8$\mu$m and 12$\mu$m, we have to take into account the uncertainty in T$_{eff}$ adopted from Table~\ref{teff}, the possible interstellar and/or circumstellar reddening of the UV and visible stellar radiation, and the fact that a part of the visible emission can originate from the circumstellar envelope (up to $\sim$ 0.5 Mag).

\begin{figure}[!t]
\centering   \includegraphics[width=0.49\textwidth]{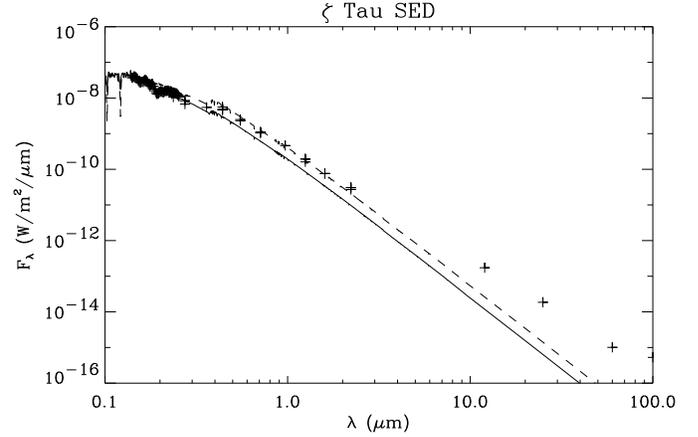}
\caption{A comparison of the reconstructed SED for $\zeta$ Tau (crosses) with two ``extreme'' models:  a reddened B2V  (solid line, T$_{\rm eff}$=22000K, R$_{\rm \star}$=5R$_{\rm \odot}$, A$_{\rm V}$=0.2, R$_{\rm V}$=3.1) for which up to 40$\%$ of the visible flux comes from the envelope, and a B4III unreddened star dominated by the stellar emission in the visible (dotted line, T$_{\rm eff}$=16000K, and  R$_{\rm \star}$=8.5R$_{\rm \odot}$).}
\label{zetSED}
\end{figure}

$\zeta$ Tau, is probably the most challenging target to model since its MK type is not well constrained: the published luminosity class ranges from II to IV and its spectral type from B2 to B4. Moreover, at 0.2$\mu$m an absorption bump is clearly evident in its UV spectra. This is a indicates the presence of interstellar and/or circumstellar extinction of the stellar emission. In Fig~\ref{zetSED} we plot the SED of $\zeta$ Tau with  two ``extreme'' models, a reddened B2V for which up to 40$\%$ of the visible flux comes from the envelope, and a B4III unreddened star fully dominated in the visible by the stellar emission. 

We obtain errors for F$_{\rm e}$ on the order of 0.06 at 8$\mu$m and 0.04 at 12$\mu$m. This simple analysis shows that the calculated N band relative envelope flux is not strongly dependent on the spectral class and associated stellar parameters, but that the main uncertainty comes from the 
envelope contribution in the visible.

We applied this method to all our seven targets to determine the envelope relative flux F$_{\rm e}$ and its uncertainty. The parameters adopted and the results of this SED model fitting are presented in Table~\ref{Fenv}. The stellar SEDs are also over-plotted in Fig~\ref{SED}.

\begin{table}[!b]
{\centering \begin{tabular}{|c|@{~}c@{~ ~}c@{~ ~}c@{~ ~}c@{~ ~}c|cc|}
\hline
		& \multicolumn{5}{c}{ Adopted stellar parameters}\vline& \multicolumn{2}{c}{F$_{\rm e}$ ($\%$)}\vline\\
Name& T$_{\rm eff}$ (K)& log g  &R$_{\rm \star}$ (R$_{\rm \odot}$)& d (pc) & A$_{\rm v}$ & 8$\mu$m &12$\mu$m\\
\hline
$\kappa$ CMa& 24600&4.07&5.9 &202&0.37 &90$\pm$6&94$\pm$4\\
$\delta$ Cen& 22360&3.92&6.5 &127&0.10 &88$\pm$5&93$\pm$3\\
$\alpha$ Ara& 18044&3.99&4.5 &81 &0.10 &87$\pm$4&91$\pm$3\\
p Car       & 17389&3.52&6.0 &148&0.05 &87$\pm$4&91$\pm$3\\
$\zeta$ Tau & 19340&3.52&7.5 &136&0.30 &80$\pm$6&86$\pm$5\\
$\alpha$ Col& 12963&3.73&5.8 &80 &0.00 &67$\pm$11&76$\pm$8\\
$\beta$ CMi & 11772&3.51&3.5 &49 &0.00 &59$\pm$13&69$\pm$10\\
\hline
\end{tabular}\par}

\caption{Stellar parameters adopted in this work and resulting 8 and 12$\mu$m envelope flux.  \label{Fenv}}
\end{table}

\begin{table*}[t]
{\centering 
\begin{tabular}{|c|c|cc|ccc|c|c|c|cc|}
\hline
&$\kappa$ CMa&\multicolumn{2}{c}{$\delta$ Cen} \vline&\multicolumn{3}{c}{$\alpha$ Ara} \vline&p Car&$\zeta$ Tau&$\alpha$ Col & \multicolumn{2}{c}{$\beta$ CMi} \vline\\
&&B$_1$&B$_2$&B$_1$&B$_2$&B$_3$&&&&B$_1$&B$_2$\\
\hline
$\phi$ (mas) at 8 $\mu$m  & $<$5.2  & 4.9 $\pm$1.8 & 4.0 $\pm$1.8 & 2.7 $\pm$1.1  & 3.9 $\pm$1.4 & 5.5 $\pm$0.3 &  $<$6.0 & $<$2.6      & $<$2.3  & $<$6.8  & $<$1.1  \\
$\phi$ (mas) at 12$\mu$m  & $<$8.4  & 5.3 $\pm$4.3 & 6.9 $\pm$2.7 & 3.6 $\pm$1.8  & 6.0 $\pm$2.3 & 8.1 $\pm$0.6 &  $<$9.5 & 5.7$\pm$2.2 & $<$3.6  & $<$10.5 & $<$1.8\\
\hline
$\phi$ (mas) at 2.1$\mu$m &$<$3.6$^2$ &\multicolumn{2}{c}{1.6$\pm$0.4 $^3$}\vline&\multicolumn{3}{c}{7.3$\pm$2 $^1$}\vline&-&1.79$\pm$0.07 $^4$&-&\multicolumn{2}{c}{-}\vline\\
\hline
\end{tabular}\par}

\caption{Envelope size from our VLTI/MIDI data and literature : Meilland et al. (2007a$^1$, 2007b$^2$, 2008$^3$) and Gies et al. (2008$^4$)}
\end{table*} 

\subsection{N band envelope extension}

We determine the N band envelope visibility, V$_{\rm e}$, using Eq.~(2) and values of F$_{\rm e}$ from Table~\ref{Fenv}. To determine the mean visibility value, V$_{\rm tot}$, at 8$\mu$m and 12$\mu$m, we use a linear interpolation of the measured visibility in the same wavelength range, assuming that the measurements at each wavelength are independent. This method allows us to slightly decrease the errors on the interpolated visibility. Finally, we determine the envelope extension using an equivalent uniform disk diameter. The results, assuming these simple models, are given in Table~\ref{extension}. For convenience, the K band measurements from VLTI/AMBER (Meilland et al. 2007a, 2007b, 2008) or CHARA (Gies et al. 2007) are also quoted if available.

As mentioned previously, the envelope is resolved only for $\alpha$ Ara, $\delta$ Cen, and $\zeta$ Tau (only at 12$\mu$m). For $\kappa$ CMa,  p Car, $\zeta$ Tau (at 8$\mu$m), $\alpha$ Col, and $\beta$ CMi, we determine a upper limit to the envelope extension only. Please see the
next section for more discussion about $\alpha$ Area's N and K band geometry, and on the compactness and wavelength dependence of the envelope size.

\section{Discussion}

\subsection{The K and N band geometry of $\alpha$ Ara}
\begin{table}[!b]
\centering\begin{tabular}{ccc}
\hline
$\lambda$($\mu$m)& major-axis (mas) & flattening\\
\hline
2.1&7.3$\pm$2&2.7$\pm$0.9\\
8  &5.8$\pm$0.8&2.4$\pm$1.0\\
12 &8.2$\pm$1.2&2.6$\pm$1.2\\					
\hline
\end{tabular}
\caption{$\alpha$ Ara equatorial disk major-axis and flattening. \label{aara_flat}}
\end{table}
First VLTI/MIDI data for $\alpha$ Ara were obtained by Chesneau et al. (2005). Unfortunately, the baselines used were too short and the data quality was not good enough to accurately constrain its geometry. The authors were only able to put an upper limit on the envelope extension, i.e. less than 10~mas at 8$\mu$m. However, this was already smaller than what Stee \& Bittar (2001) predicted for this Be star. Thus, these authors proposed a possible disk truncation and found spectroscopic evidence for the presence of a companion orbiting outside the estimated outer-radius of the equatorial disk, confining the disk within the Roche lobe of the star.

Using VLTI/AMBER data, Meilland et al. (2007a) investigated $\alpha$ Ara K-band geometry and kinematics. They clearly resolved the equatorial disk and measured its flattening and orientation. The disk orientation was in agreement with the value deduced from polarimetric measurements by Yudin \& Evans (1998) (P.A. =166$^o$). Meilland et al. (2007a) also found that a few percent of the K band flux originates from an extended structure in the polar direction. Using the SIMECA code developed by Stee et al. (1996), they were able to model this object assuming a quasi-critical rotator surrounded by both a dense Keplerian rotating disk and an enhanced polar wind with terminal expansion velocities up to 1000 kms$^{-1}$. Note that this is the first time that the nature of the disk rotation, found to be Keplerian, was directly measured for a Be star.

Due to the SCI-PHOT mode and the use of longer baselines (i.e. UT1-UT4), the quality of our new VLTI/MIDI data was good enough to accurately constrain the envelope geometry in the N band. We were able to successfully model our measurements using a uniform ellipse, and found the same value for the major-axis as determined in the K-band from the VLTI/AMBER data by Meilland et al. (2007a). The envelope extension (a) and flattening (the ratio of a/b) are given in 
Table~\ref{aara_flat}.

\begin{figure}[!b]
 \centering\includegraphics[width=0.48\textwidth]{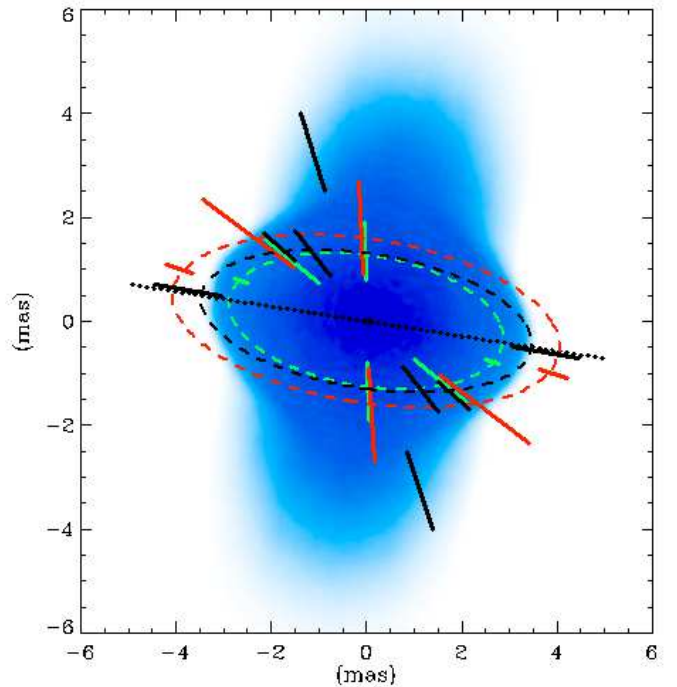}
\caption{$\alpha$~Ara envelope at 8$\mu$m (green bars) and 12$\mu$m (red bars) assuming a uniform disk + an unresolved star model plotted as a function of baseline orientation.  VLTI/AMBER measurements (black bars) are from Meilland et al. (2007a) and the corresponding best SIMECA model in the continuum ( shaded background ) are also shown. The black, green, and red ellipses correspond to the uniform ellipse fit of the equatorial disk presented in Table~6 at 2.1, 8, and 12$\mu$m, respectively. The orientation of the major axis (dotted black line) is perpendicular to polarization angle measurement obtained by Yudin \& Evans (1998).}
\label{alpha_ara}
\end{figure}

 Both the VLTI AMBER and VLTI/MIDI measurements produced similar disk sizes over the wavelengths, 2.1$\mu$m, 8$\mu$m, and 12 $\mu$m. The envelope extension is plotted as a function of the baseline orientations in Fig~\ref{alpha_ara}. A SIMECA image in the K band from Meilland et al. (2007a) is also plotted in this figure to allow a direct comparison with the K and N band measurements. 

\begin{figure*}[!t]
\centering   \includegraphics[width=0.92\textwidth]{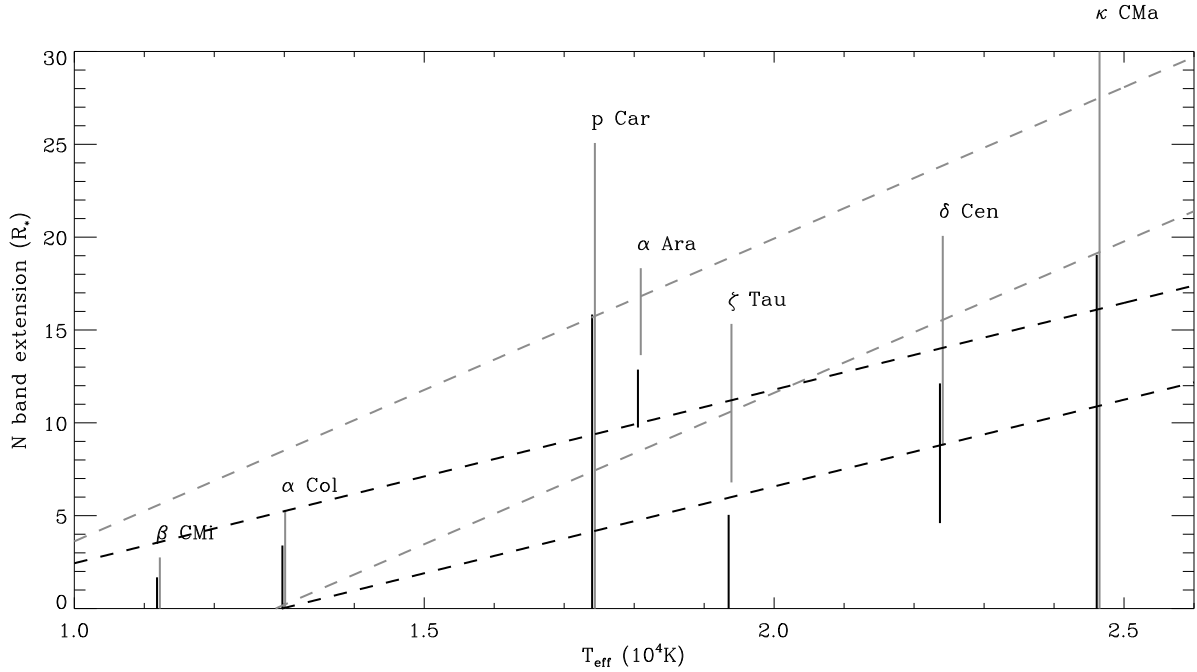}
\caption{ 8$\mu$m (black) and 12$\mu$m (gray) envelope extension  as a function of T$_{\rm eff}$ for the program Be stars. The dotted lines represent linear interpolations, accounting for the uncertainty of the measurements, of the temperature dependence on the N band size. }
\label{taille}
\end{figure*}

One of the most striking results from our modeling is the fact that the 
equatorial disk extensions remain relatively constant between 2.1$\mu$m and 
12$\mu$m.  
We originally expected that longer wavelengths would probe cooler regions and thus larger disk volumes, which is clearly not the case.
However, small variations on the order of $\sim 2\sigma$ are detected between 8$\mu$m and 12$\mu$m along the direction of the major-axis. This wavelength independent disk size may be evidence for disk truncation by a companion, in agreement with the suggestion of Meilland et al. (2007a). However, a free-free and 
free-bound radiation dominated envelope with a non-truncated isothermal equatorial disk may also give the same result depending on thermal structure of the 
disk. For example, studies that have determined self-consistent disk 
temperature distributions based on the assumption of radiative equilibrium,
find that, on average, the temperature distribution is nearly constant up to 
$\sim$ 100 R$_*$ and is approximatively half the T$_{\rm eff}$ of the central 
star.  However, depending on the equatorial density near the stellar 
surface, large variations in temperature can occur especially in dense disks.
For example, see Millar \& Marlborough (1998) for thermal structure of pure hydrogen disks, 
Jones et al. (\cite{jones}) for the effects of iron line cooling on temperatures, or Sigut \& Jones (\cite{sigut}) for models that include a realistic solar
type chemical composition.  Thus, it may be difficult to propose a general 
scheme for each of our program Be stars since the disk density may vary 
significantly from star to star.
 
Another interesting result can be seen from Fig~\ref{alpha_ara}. An apparent lack of an extended polar contribution in the N band data compared to the K band measurement is apparent. There are two possible explanations:
\begin{enumerate}
\item The VLTI/MIDI angular resolution is 4-6 times smaller than the VLTI/AMBER. The polar wind extends out to $\sim$ 6~mas and is almost fully resolved by VLTI/AMBER with 80m baselines, whereas, the VLTI/MIDI visibilities with 120m baselines are $\sim$ 0.8. Thus, the overall effect on the total visibility for the VLTI/MIDI instrument is reduced by a factor of $\sim$5.\\

\item If the poles are hotter than the equatorial regions because of the effects of gravity darkening, the relative contribution to the total flux of the polar wind can significantly decrease between 2.1 and 8$\mu$m. In order to estimate this effect, a full modeling of the star+disk+wind would be required which is well beyond the scope of this paper.
\end{enumerate}

\subsection{The geometry of the remaining program Be stars}

\begin{table*}[!t]
{\centering \begin{tabular}{|c|c|cc|ccc|c|c|c|cc|}
\hline
&$\kappa$ CMa&\multicolumn{2}{c}{$\delta$ Cen}\vline &\multicolumn{3}{c}{$\alpha$ Ara}\vline &p Car&$\zeta$ Tau&$\alpha$ Col & \multicolumn{2}{c}{$\beta$ CMi}\vline \\
\hline
Polarization($\%$)          &0.3&\multicolumn{2}{c}{0.33}\vline&\multicolumn{3}{c}{0.58}\vline&1.16&1.13&0.15&\multicolumn{2}{c}{0.14}\vline\\
Polarization angle ($^o$)                  &106&\multicolumn{2}{c}{137}\vline &\multicolumn{3}{c}{166}\vline &68  &34  &109 &\multicolumn{2}{c}{90}\vline  \\
\hline
Baseline orientation ($^o$) &&B$_1$&B$_2$&B$_1$&B$_2$&B$_3$&&&&B$_1$&B$_2$\\
compared to the disk major-axis &9&4.5&19.3&59&21.6&1.9&62.8&56.1&45&63.8&64.5\\
\hline
\end{tabular}\par}
\caption{Yudin \& Evans (1998) polarization measurements and baselines orientation with respect to the disk putative major-axis.\label{pola}}
\end{table*}


As mentioned above, the envelopes of all observed targets are much smaller than predicted by Stee \& Bittar (2001). In fact, none of our disks are larger than 10~mas. Using the angular diameters from Table 3 and parameters from Table 2, each disk is plotted (in R$_\star$) in the N band as a function of the central stars' T$_{\rm eff}$ in Fig~\ref{taille}. This figure seems to indicate that the envelopes of late type Be stars ($\alpha$ Col and $\beta$ CMi) are smaller than those of earlier types. However, this claim should be taken with caution considering the uncertainty of the measurements.

Moreover, with the exception of $\alpha$ Ara, one has to keep in mind that these stars are measured with baselines not necessarily aligned with their major-axis. Comparing polarization measurements of Yudin \& Evans (1998) (see Table~\ref{pola}) and the baselines orientations from Table 1, it follows that the baseline orientations of $\alpha$ Col, $\beta$ CMi, $\zeta$ Tau, and p Car correspond to an intermediate angle between their putative major and minor axis whereas $\delta$~Cen and $\kappa$ CMa are certainly observed along their major axis. 
 
Three stars in our sample were also observed with VLTI/AMBER~: $\alpha$ Ara, $\kappa$ CMa, $\delta$ Cen and their corresponding extensions in the K band continuum are 7.3$\pm$2, $<$3.6, and 1.6$\pm$0.4~mas, respectively. $\zeta$ Tau was observed with the CHARA interferometer in the K$^\prime$ band (Gies et al. 2007) and modeled with an elliptical Gaussian with a FWHM extension of 1.79$\pm$0.07~mas. 

For $\kappa$ CMa, the N band measurements are not accurate enough to determine if its envelope size varies with wavelength. On the other hand, $\delta$ Cen and $\zeta$ Tau seem to show some size variations between the K and N bands. For $\zeta$ Tau, Carciofi et al. (2009) have shown in a recent paper that a density wave seems to be orbiting within the disk. Thus, the apparent extension of the disk may be different at different times.As previously discussed, this is contrary to the findings for $\alpha$ Ara as 
we showed that its circumstellar disk extension is relatively constant over the infrared wavelengths considered. 

Overall, for the Be star systems considered here, the variations in disk size between the K and N bands seem relatively small.  This is quite different from observations in these bands for the dusty disks of supergiant B[e] stars. For example, Domiciano de Souza et al. (2007) find that the interferometric observations of the circumstellar environment for the B[e] star, CPD-57$^o$2874 in the K and N band are well fitted by an elliptical Gaussian model with FWHM diameters that varies linearly with wavelength. However, the N-band emission mechanisms are different for these two classes of stars, i.e. free-free emission for classical Be stars and optically thick thermal dust emission for B[e] stars. Nevertheless, comparing K and N band emissions of circumstellar envelopes of hot stars may provide clues, allowing the physics of these systems to be deciphered.

\section{Conclusion}

This first mid-infrared interferometric statistical study of classical Be 
stars shows evidence for the compactness of their circumstellar environment in the N band, smaller than 10 mas for the all Be stars in our sample. Interestingly, Gies et al. (2007) find that Be stars in binary systems generally have smaller disks. 

Further, Be star disk extensions do not seem to depend strongly on the observed wavelength between 8 $\mu$m and 12 $\mu$m. For $\alpha$ Ara, an even stronger case can be made since the disk exhibits the same extension and flattening in the K and N bands. A possible result, that needs further investigation, is whether or not the disk size in various bands correlates with the  T$_{\rm eff}$ of the central star. Our investigation seems to suggest that there may be such a correlation, but
this must be taken with caution considering the large uncertainties in the data and the fact that, in most cases, the baselines were not aligned with the disks' 
major axis.

This study can only be marginally extended to a larger number of targets because of VLTI/MIDI limited sensibility and accuracy. Fortunately, the number of targets available and the precision will significantly increase with MATISSE (Lopez et al. 2008), the second generation VTLI mid-infrared beam-combiner.

\begin{acknowledgements} 
The Programme National de Physique Stellaire (PNPS) and the Institut National en Sciences de l'Univers (INSU) are acknowledged for their financial support.  A. Meilland acknowledges financial support from the Max Planck Institut fur Radioastronomy, and thanks Thomas Driebe for his help in reducing MIDI data. CEJ wishes to acknowledge financial support from NSERC, Natural Sciences and Engineering Research Council of Canada. The authors would like to thank the referee, Thomas Rivinius, for his careful reading and suggestions that greatly improved the paper. This research has made use of SIMBAD and VIZIER databases, operated at CDS, Strasbourg, France.
\end{acknowledgements}

\end{document}